\documentclass[a4paper,11pt]{article}
\pdfoutput=1 % if your are submitting a pdflatex (i.e. if you have
\usepackage{rotating}
\usepackage{jinstpub} % for details on the use of the package, please
\title{\boldmath Design and first tests of the Gotthard-II readout ASIC for the European X-ray Free-Electron Laser}

\author[a,1]{Jiaguo Zhang,\note{Corresponding author.}}
\author[a]{Marie Andr\"a,}
\author[a]{Rebecca Barten,}
\author[a]{Anna Bergamaschi,}
\author[a]{Martin Br\"uckner,}
\author[a]{Sabina Chiriotti-Alvarez,}
\author[a]{Roberto Dinapoli,}
\author[a]{Erik Fr\"ojdh,}
\author[a]{Dominic Greiffenberg,}
\author[a]{Pawel Kozlowski,}
\author[b]{Markus Kuster,}
\author[a]{Carlos Lopez-Cuenca,}
\author[a]{Markus Meyer,}
\author[a]{Davide Mezza,}
\author[a]{Aldo Mozzanica,}
\author[b]{Marco Ramilli,}
\author[a]{Christian Ruder,}
\author[a]{Bernd Schmitt,}
\author[a]{Xintian Shi,}
\author[a]{Dhanya Thattil,}
\author[a]{Gemma Tinti,}
\author[b]{Monica Turcato,}
\author[a]{Seraphin Vetter}

\affiliation[a]{Paul Scherrer Institut,\\5232 Villigen, Switzerland}
\affiliation[b]{European XFEL,\\Holzkoppel 4, 22869 Schenefeld, Germany}

\emailAdd{jiaguo.zhang@psi.ch}

\abstract{Gotthard-II is a charge-integrating microstrip detector developed for experiments and diagnostics at free-electron lasers using hard X-rays of 5 keV - 20 keV. Thanks to its excellent single photon sensitivity, large dynamic range as well as high frame rate of 4.5 MHz in burst mode, its potential scientific applications include X-ray absorption/emission spectroscopy, hard X-ray high resolution single-shot spectrometry (HiREX), beam diagnostics, as well as veto signal generation for pixel detectors. The Gotthard-II ASIC has been designed and fabricated using \mbox{UMC-110 nm} technology. The final ASIC design and performance in terms of noise, linearity, dynamic range, coupling between channels and speed will be discussed in the paper. In addition, a first measurement of an X-ray absorption spectrum of a standard copper sample has been done. The performance of the Gotthard-II in an experiment using energy dispersive X-rays has been demonstrated.
}


\begin{document}
\maketitle
\flushbottom

%\linenumbers

\section{Introduction}
\label{sec:intro}

Gotthard-II is a charge-integrating microstrip detector developed for hard X-ray experiments at Free-Electron Lasers (FELs), in particular for the European X-ray Free-Electron Laser (EuXFEL) in Schenefeld \cite{Altarelli2006, Decking2020, Tschentscher2017}. The EuXFEL delivers ultrashort (\mbox{$<$ 100 fs}), high intensity ($10^{12}$ ph/pulse) and fully coherent X-ray pulses with a peak brilliance $\sim$8 orders of magnitude higher than any synchrotron radiation source. Photon-counting detectors cannot be used at FELs due to simultaneously impinging  X-ray photons. Instead, charge-integrating detectors must be used and are considered as the only option for FEL applications. Among all FELs, the EuXFEL has a unique bunch structure: a bunch spacing of 220 ns in-between the 2700 pulses of a train, with a train repetition rate of 10 Hz. It poses great challenges to detectors in terms of \mbox{4.5 MHz} frame rate and requires a local storage of images in the ASIC since it is unfeasible to readout images during the 220 ns bunch spacing.

Among all detectors for the EuXFEL, Gotthard-II will be the most widely employed detector for energy dispersive experiments well suited to the 1-D geometry. The Gotthard-II detector uses a silicon microstrip sensor with a pitch of 50 $\mu$m or 25 $\mu$m and with 1280 or 2560 channels wire-bonded to 10 or 20 readout ASICs respectively. The scientific and operation requirements at the EuXFEL are very challenging, especially in the view of the development of the Gotthard-II readout ASIC. %To fulfil the requirements at the EuXFEL, it is very challenging in terms of the development of the Gotthard-II readout ASIC. 

The Gotthard-II readout ASIC features: (a) a high speed, dynamic gain switching pre-amplifier (PRE) allowing to cope with the 4.5 MHz frame rate. The implementation of the dynamic gain switching circuit enables the detection of up to $10^{4}$ X-ray photons of 12.4 keV maintaining single photon resolution with a Signal-to-Noise Ratio (SNR) greater than 10 for low photon fluxes; (b) an on-chip Analog-to-Digital Converter (ADC) and a Static  Random-Access Memory (SRAM) capable of storing all digitized images from the 2700 pulses in a bunch train; (c) an on-chip digital comparison circuit to generate veto signals for pixel detectors, for example AGIPD~\cite{Aschkan2019JSR} and LPD~\cite{Hart2012}, which are able to record at most 352 and 512 images per bunch train, respectively. With the veto signals, memories of the pixel detectors storing empty images due to poor interactions between X-ray pulses and samples in user experiments or due to unqualified X-ray pulses can be re-used for the other forthcoming pulses in the same bunch train. The detailed specifications and development strategy of Gotthard-II for the EuXFEL can be found in \cite{Zhang2018, Monica2014}. In addition, the Gotthard-II ASIC is capable of taking images continuously at a frame rate of less than 495 kHz for synchrotron experiments, the future upgrade of the EuXFEL to CW mode, and potentially for other FELs which are planned to be operated in CW mode, for example LCLS-II~\cite{LCLS} and SHiNE~\cite{SHINE}. 

In this paper, the architecture of the ASIC will be explained in detail. The performance of the final ASIC design in terms of noise, linearity, dynamic range, coupling between channels as well as signal settling time have been investigated, and their results will be discussed.

\section{The Gotthard-II ASIC design}

\begin{sidewaysfigure}
\small
\centering
\includegraphics[width=22cm]{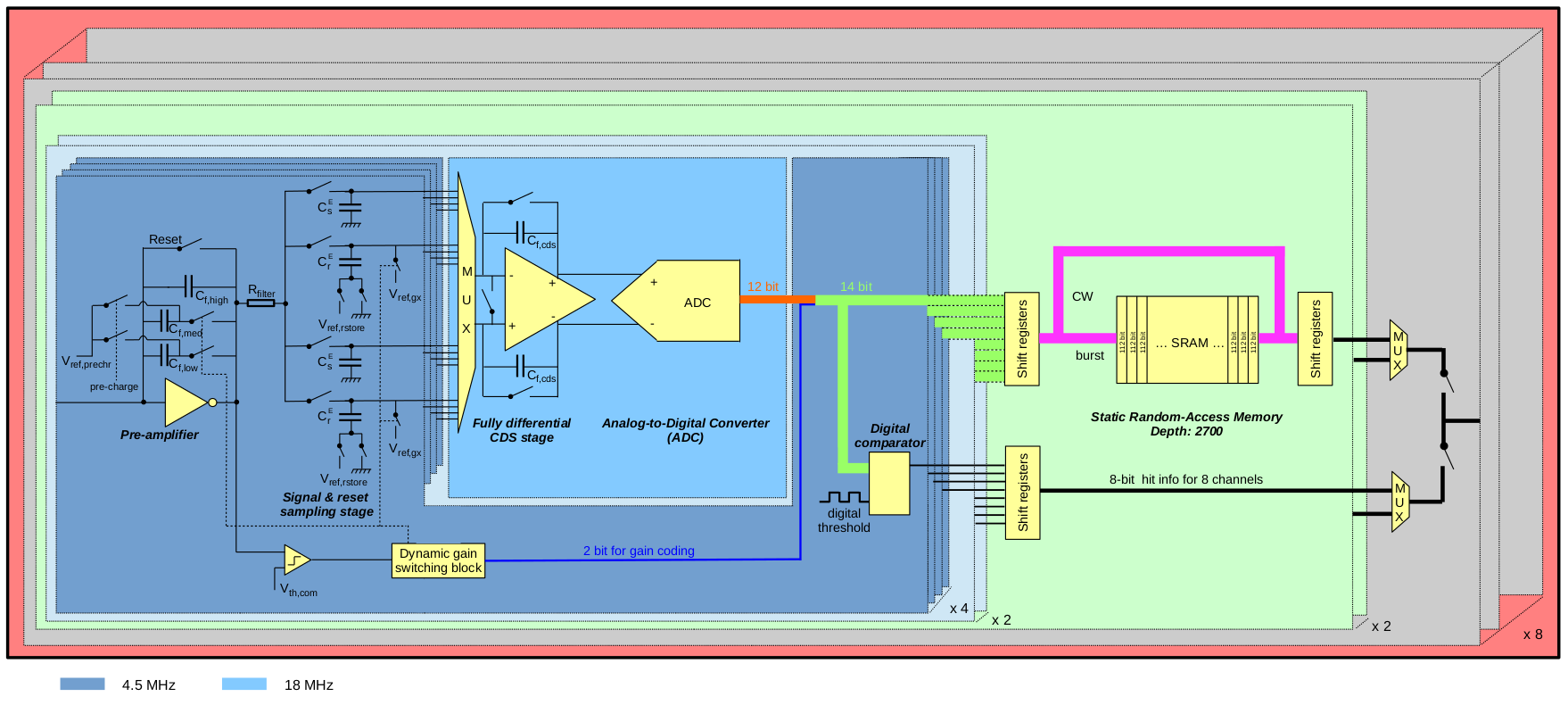}
\caption{The architecture of the Gotthard-II readout ASIC. For usage at the EuXFEL, the preamplifier and digital comparator run at 4.5 MHz, while the correlated-double-sampling stage and analog-to-digital converter run at 18 MHz.}
\label{Architecture}
\end{sidewaysfigure}

The Gotthard-II readout ASIC has been designed and fabricated using \mbox{UMC-110 nm} technology~\footnote{110 nm, aluminum only, mixed mode UMC110AE technology node with analog options~\cite{UMC110}.}. The architecture of the ASIC is shown in figure~\ref{Architecture}: it includes 128 dynamic gain switching pre-amplifiers (PRE) and digital comparators, 32 fully differential Correlated-Double-Sampling (CDS) amplifiers and Analog-to-Digital Converters (ADC) with 12 bits, as well as 16 Static Random-Access Memories (SRAM) with a depth for 2720 images. 

\begin{figure}[hbt!]
\small
\centering
\includegraphics[width=14.5cm]{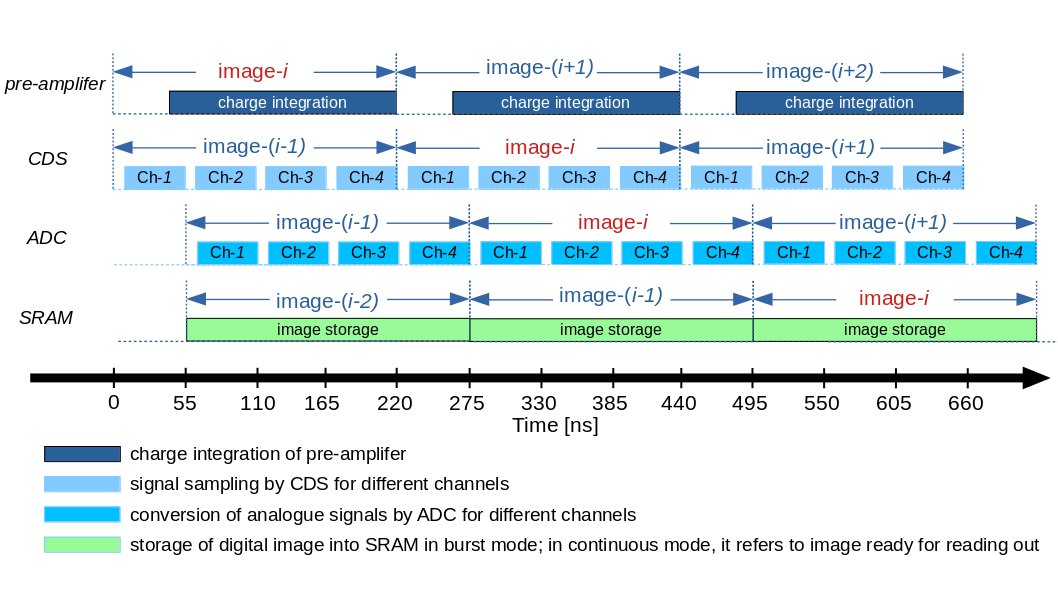}
\caption{The image processing pipeline in the Gotthard-II readout ASIC. }
\label{Pipeline} 
\end{figure}

%Signals in the Gotthard-II readout ASIC are processed in a pipeline.The pipeline processing is shown in figure~\ref{Pipeline}: During the charge integration of the PRE for image-$i$ (time slot 0 - \mbox{220 ns}), the CDS samples the analogue signals from image-($i-1$) and the ADC converts these signals into digital ones with a phase shift of 90 degree (55 - 275 ns). While the ADC starts the conversion for image-($i-1$), image-($i-2$) is stored into the SRAM (55 - 275 ns) or available for reading out directly depending on the ASIC configuration. 
Signals in the Gotthard-II readout ASIC are processed in a pipeline. The pipeline processing is shown in figure~\ref{Pipeline}: First, image-$i$ is processed by the PRE from 0 to 220 ns. Then the CDS samples the analogue signals at the output of the PRE for image-$i$ from 220 to 440 ns and the ADC digitizes the signals with a phase shift of 90 degree (275 - 495 ns). After 495 ns, the digital signals from the ADC output are stored into the SRAM or available for reading out directly depending on the ASIC configuration.

The ASIC can be configured to either burst mode for the EuXFEL or continuous mode for synchrotron applications:

1) In burst mode, images are continuously taken and stored into different addresses of the SRAM and then read out during the bunch train spacing of 99.4 ms. The PRE and digital comparator of each channel run at 4.5 MHz, while the CDS and ADC shared by every four neighbouring channels run at a sampling and conversion rate of \mbox{18 MHz} (namely 18 MSamples/s or 18 MS/s). The output of the digital comparator with 1 bit per channel storing the hit information is read out every \mbox{220 ns} with a latency of 495 ns due to the pipeline architecture of the ASIC, which will be explained in the following sections.

2) In continuous mode, images are continuously taken and read out without storing them into different addresses of the SRAM. The PRE runs at the same frequency as the frame rate, while the CDS and ADC at four times the frame rate. The minimal readout time of each frame is $\sim$ 2 $\mu$s.

%The signals in the ASIC are processed in a pipeline architecture: When the signal generated by bunch-$i$ is being integrated by the PRE and stored into one set of the analogue storage cells in the "signal \& reset sampling stage", the signal generated by the previous one (bunch-($i-1$)) stored in the other set of analogue storage cells is being "processed" by the CDS and ADC, while at the same time the output of the ADC after converting the signal generated by bunch-($i-2$) is being stored into the SRAM. The PRE and digital comparator of each channel runs at 4.5 MHz, while the CDS and ADC at \mbox{18 MHz} sampling rate, since one ADC is shared between 4 channels. The images saved in SRAM are readout in-between the bunch train spacing of the XFEL.EU of 99.4 ms. The output of the digital comparator as the veto signal is readout every \mbox{220 ns} from the readout chip.

\subsection{Pre-amplifier}

The PRE of the ASIC is a charge-sensitive pre-amplifier with dynamic gain switching functionality. Although it is similar to \mbox{Gotthard-I~\cite{Aldo2011}}, AGIPD~\cite{Xintian2012, Julian2012, Aschkan2015, Aschkan2019} and JUNGFRAU~\cite{Aldo2014}, the PRE of the Gotthard-II ASIC has further optimizations for its DC gain in order to reduce the coupling effect, also known as "cross talk". A high DC gain design is of importance when using silicon sensors with large inter-electrode capacitance. For a PRE with small DC gain, even if all carriers generated by X-ray photons are collected by one strip, its neighbouring strip channels can still measure a certain amount of charge due to the charge division between neighbouring channels through the inter-electrode capacitance. This makes the calibration rather complex as already discussed in ~\cite{Zhang2017,Zhang2018}. Instead of using a cascaded push-pull inverter which was implemented in the Gotthard-I ASIC, the PRE of Gotthard-II is based on a split-transistor push-pull inverter, which provides a high DC gain up to 600-900 for working voltages from 1.2 V to 1.4 V as well as a low power consumption down to 0.25 mW per channel~\cite{Zhang2017}. The PRE runs at 4.5 MHz for the EuXFEL.

In the PRE, three feedback capacitors with different capacitance values are implemented: $C_{f,high}$, $C_{f,med}$ and $C_{f,low}$ with increasing capacitance. The output of the PRE is connected to a comparator followed by a dynamic gain switching logic block. Initially, $C_{f,high}$, the smallest of the three capacitances, is used as the feedback capacitor of the PRE, which provides the highest gain, lowest noise as well as single photon resolution. During integration (after the "Reset" switch is released), the output voltage of the PRE moves according to the total charge integrated on the feedback capacitor: $\Delta V_{PRE} = - \Delta Q/C_{f,high}$. When the output voltage crosses the threshold of the comparator, $V_{th,com}$, the dynamic gain switching block will force the gain switching of the PRE: the second capacitor $C_{f,med}$ will automatically be added to the feedback loop, which results in an increased feedback capacitance and thus a reduction of the PRE output. Thus, the change in voltage at the PRE output is given by $\Delta V_{PRE} = - \Delta Q/(C_{f,high}+C_{f,med})$, $i.e.$ the pre-amplifier gain after the charge re-distribution on the two feedback capacitors is reduced. After the first gain switching, if the PRE output is still above $V_{th,com}$, a second gain switching occurs by adding the largest capacitor $C_{f,low}$ to the circuit which further reduces the PRE output. In this case, the PRE output is given by $\Delta V_{PRE} = - \Delta Q/(C_{f,high}+C_{f,med}+C_{f,low})$. The gain used by the PRE is coded using 2 digital bits, which are named gain bits. The three gains are called high, medium and low gain respectively and expressed simply as G0, G1 and G2.

The capacitance values of $C_{f,high}$, $C_{f,med}$ and $C_{f,low}$ are carefully selected so that both single photon resolution at low flux as well as large dynamic range up to \mbox{$10^{4} \times 12.4$ keV} at high flux can be achieved. In addition, the two capacitors, $C_{f,med}$ and $C_{f,low}$, can be pre-charged using an external voltage, $V_{ref,prech}$, during the PRE reset phase (when the "Reset" switch is closed). Right after the gain switching occurs, the pre-stored charge on $C_{f,med}$ or $C_{f,low}$ will be added to the feedback loop, which moves the offset of the PRE output in the second and third gain. Thus, by adjusting $V_{ref,prech}$ the dynamic range can be maximized.

The output of the PRE can be connected to two sets of analogue storage cells (each noted as $"E"$ for even images or $"O"$ for odd images in the "Signal \& reset sampling stage" in figure~\ref{Architecture}) through a resistor, $R_{filt}$. Each set of the analogue storage cells includes two storage capacitors with the same capacitance value: $C_{r}^{E/O}$ and $C_{s}^{E/O}$. $C_{r}^{E/O}$ is used to record the output of the PRE after the "Reset" switch is released and before X-rays impinging, which is useful to remove the low frequency noise of the PRE and the additional charge due to the "Reset" switch; while $C_{s}^{E/O}$ stores the additional signal induced by the X-ray photons. The resistor $R_{filt}$ connecting to the PRE output filters high frequency noise in connection with $C_{r}^{E/O}$. The resistance of $R_{filt}$ can be set through two configurable bits, which results in a value of 0, 10, 50 or 400 k$\Omega$. 
%Signals on the two analogue storage cells are subtracted and amplified by the differential CDS stage in the subsequent circuit.

\subsection{Correlated-Double-Sampling stage}

Every four channels are connected to one fully differential CDS stage through a multiplexer (MUX). The CDS stage multiplexes at a rate of 18 MHz sampling rate, namely four times faster than the PRE. The CDS stage is a switched capacitor amplifier implemented with a fully differential folded-cascode operational transconductance amplifier and a switched capacitor common mode feedback circuitry. 

%Before gain switching,  $C_{r}^{E/O}$ is connected to $V_{ref,rstore}$ during sampling the PRE output and to ground when $C_{s}^{E/O}$ is sampling the PRE output during X-ray photons impinging. 
While sampling the reset signal the bottom plate of the capacitor, $C_{r}^{E/O}$, is connected to $V_{ref,rstore}$. During acquisition, it is connected to ground (see figure~\ref{Architecture}). This effectively works like a precharge to maximize the dynamic range.  The fully differential CDS amplifier converts the single-ended signals to differential ones. The difference of the output differential signals is given by:

\begin{equation}
\Delta V_{out} = V_{out}^{+} - V_{out}^{-} = [V_{s}^{E/O} - (V_{r}^{E/O} - V_{ref,rstore})] \times C_{i}/C_{f} 
\label{eq1}
\end{equation}

\noindent where $C_{i} = C_{r}^{E/O} = C_{s}^{E/O}$ is the value of the capacitance of the analogue storage cells and $C_{f}$ the feedback capacitance of the CDS amplifier. $V_{out}^{+}$ and $V_{out}^{-}$ are the differential voltages at the CDS output.

After gain switching, a fixed voltage $V_{ref,gx}$ is connected to one input of the differential CDS stage. In this case, the CDS output can be written as: 

\begin{equation}
\Delta V_{out} = V_{out}^{+} - V_{out}^{-} = (V_{s}^{E/O} - V_{ref,gx}) \times C_{i}/C_{f}
\label{eq2}
\end{equation}

The differential output of the CDS stage is centered at a common mode voltage $V_{cm}$. Thus, the differential signals are given by:

\begin{equation}
V_{out}^{+} = V_{cm} + \Delta V_{out}/2   
\label{eq3}
\end{equation}

\begin{equation}
V_{out}^{-} = V_{cm} - \Delta V_{out}/2   
\label{eq4}
\end{equation}

The common mode voltage $V_{cm}$ is typically set to \mbox{0.7 V} for Gotthard-II. By adjusting $V_{ref,rstore}$ in high gain or $V_{ref,gx}$ in medium and low gain, the differential output of the CDS stage can be adapted to the input range of the ADC for the entire dynamic range of the Gotthard-II. 

The gain of the CDS amplifier is determined by $C_{i}/C_{f}$, which can be configured to either 2.2 or 3.0 by adjusting the value of $C_{f}$. The high gain setting of the CDS amplifier shows better noise performance than the low gain setting; however, it reduces the dynamic range of the detector as a drawback. Where not otherwise mentioned, low gain setting for the CDS amplifier is used during the characterization.

\subsection{Analog-to-Digital Converter}

The differential output of the CDS stage is connected to an ADC. The ADC implemented into the Gotthard-II ASIC is a fully differential asynchronous Successive Approximation Register (SAR) ADC. This type of ADC is well-known for its low power consumption and in particular it does not require a high frequency clock for each comparison compared to a synchronous design. The ADC in the Gotthard-II ASIC provides 12-bit raw output with a sampling and conversion rate greater than 18 MS/s adapted to the EuXFEL. Although the target resolution is 10-bit, the additional 2 bits in the design allow for a calibration of the ADC for its missing codes and non-linearity, etc.

\begin{figure}[hbt!]
\small
\centering
\includegraphics[width=12.5cm]{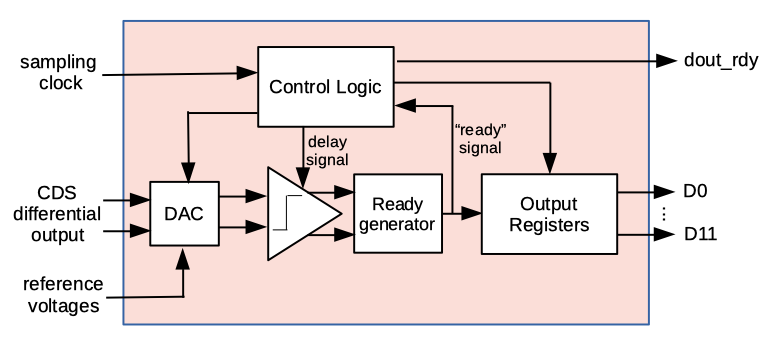}
\caption{The block diagram of the SAR ADC in the Gotthard-II readout ASIC. }
\label{ADC_logic} 
\end{figure}

The block diagram of the ADC is shown in figure~\ref{ADC_logic}. It includes: 1) a charge redistributed capacitive Digital-to-Analog Converter (DAC) array with two banks~\cite{McCreary1975}, 2) a comparator, 3) a "ready" signal generator, 4) a control logic, as well as 5) output registers. The ADC has only one comparator, thus it takes 12 comparison steps to generate the 12-bit raw output. At the beginning of each comparison cycle, the first comparison step for the Most Significant Bit (MSB) decision is executed right after the input is sampled. The other 11 comparison steps in the rest are asynchronously triggered by the "ready" signal generated by the "Ready generator" after the comparator completes each comparison step~\cite{Chen2006}.

\begin{figure}[hbt!]
\small
\centering
\includegraphics[width=14.5cm]{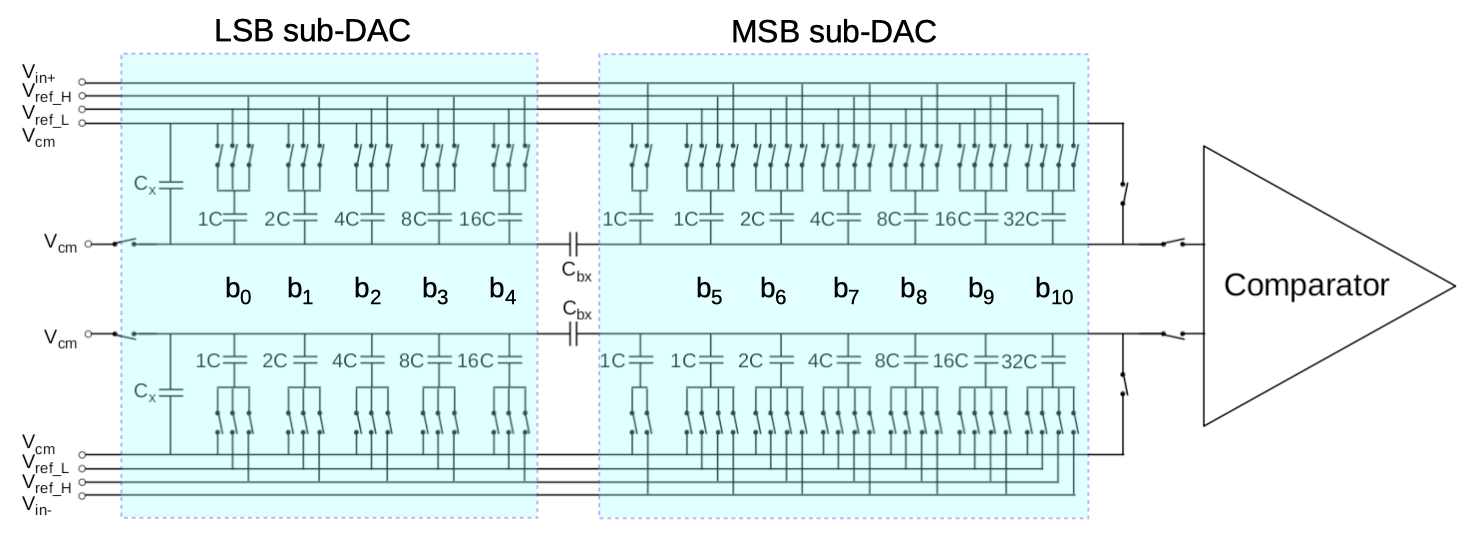}
\caption{The architecture of one of the DAC arrays of the ADC. Both DAC banks are identical in the design and thus only one is shown. Figure is reproduced from ~\cite{Zhang2018}.}
\label{ADC_DAC} 
\end{figure}

Different from a conventional DAC array using capacitors with an exponential increase in its value with a base of 2, the design for the Gotthard-II ADC is based on a split binary-weighted capacitor DAC, reducing the power consumption as well as the total capacitance, which enables less charging time and thus a higher speed~\cite{Baker1998}. The architecture of the DAC array is shown in Fig~\ref{ADC_DAC}.

Two DAC banks are implemented in the ADC so that simultaneous sampling and comparison can be done with no dead time: When one DAC bank is sampling the CDS output, the other one is connected to the ADC comparator for comparisons. 

Each DAC bank consists of two sub-DAC arrays at each input of the comparator: LSB sub-DAC and MSB sub-DAC. The two sub-DAC arrays are separated by a bridge capacitor, $C_{bx}$, whose capacitance is equal to a unit capacitance of $C$ in figure~\ref{ADC_DAC}. In addition, a side capacitor labeled as $C_{x}$ is implemented in the LSB sub-DAC. Due to the existence of a parasitic capacitance parallel to $C_{bx}$ (typically $C_{bx} > 1C$ after fabrication), it is necessary to match the capacitance of the equivalent circuit consisting of the LSB sub-DAC in serial to $C_{bx}$ with the capacitance of the smallest capacitor in the MSB sub-DAC by tuning the capacitance of $C_{x}$. Thus, a better linearity from the ADC output as well as code continuity can be achieved. The parasitic capacitance parallel to $C_{bx}$ from post-layout simulation is found to be $\sim$ 19.4 fF. To match the fore-mentioned capacitance values, $C_{x}$ can be set with 4 configurable bits ranging from 400 fF to 1 pF. 

During the sampling phase, one of the two DAC banks is disconnected from the comparator and it is charged up by the input signals, $V_{in,+}$ and $V_{in,-}$, while keeping both sides of both $C_{bx}$ connected to the common-mode voltage, $V_{cm}$. $V_{in,+}$ and $V_{in,-}$ are identical to the differential voltages from the CDS output, $V_{out}^{+}$ and $V_{out}^{-}$, as discussed in the prevous section. During the comparison phase, the DAC bank is disconnected from the input signals but connected to $V_{cm}$ in the meanwhile leaving the nodes at the two sides of $C_{bx}$ floating; thus, the input signals of the ADC are presented to the input nodes of the comparator $V_{comp,+}$ and $V_{comp,-}$, with $V_{comp,+} = 2V_{cm} - V_{in,+}$ and $V_{comp,-} = 2V_{cm} - V_{in,-}$ respectively. To allow the signal settling at the input of the comparator, a delay can be configured through 3 bits before each comparison is made. After the delay time ends, the first comparison is made directly between the voltages at the two inputs of the comparator. The comparison result for bit-11 ($b_{11}$, MSB) is generated and stored into the output registers. In additon, a "ready" signal is generated through the "Ready generator" and sent to the control logic of the ADC, which controls the switches in the DAC array and sends the "delay signal" to the comparator before the next comparison is made. If the voltage at the "+" input of the comparator is smaller than its "-" input, namely $V_{comp,+} < V_{comp,-}$, the capacitor for $b_{10}$ at the "+" input side is connected to the high reference voltage, $V_{ref,H}$, while the capacitor at the "-" input to the low reference voltage, $V_{ref,L}$: For an ideal DAC without capacitance mismatch and parasitics, this increases $V_{comp,+}$ by $|V_{ref,H}-V_{cm}|/2$ and reduces $V_{comp,-}$ by $|V_{cm}-V_{ref,L}|/2$. Similarly, if $V_{comp,+} > V_{comp,-}$, $V_{comp,+}$ is reduced by $|V_{cm}-V_{ref,L}|/2$ and $V_{comp,+}$ increased by $|V_{ref,H}-V_{cm}|/2$. Since the settings for $V_{ref,H}$ and $V_{ref,L}$ are usually symmetric and centered at $V_{cm}$, the changes of voltage at both inputs of the comparator are identical and given by $|V_{ref,H}-V_{cm}|/2 = |V_{cm}-V_{ref,L}|/2$ \footnote{The ADC can tolerate a deviation from this condition, but the mathematical treatment of the general case is out the scope of the paper}. If there is no capacitance mismatch in the DAC array, after each comparison, the input signals of the comparator change by: $|V_{ref,H}-V_{cm}|/2^{i}$, with $i$ the $i$-th comparison. 

The power consumption of the ADC, which is determined from an independent ADC test structure with the same design, is $\sim$ 1 mW at 18 MS/s sampling rate. Thus for the Gotthard-II readout ASIC, consisting of 32 ADCs, the power consumption from all ADCs is $\sim$ 32 mW when running in burst mode at the EuXFEL.

%\begin{center}
%\tabcaption{ \label{ADC_parasitic}  The configurable parasitic capacitance $C_{x}$.}
%\footnotesize
%\begin{tabular*}{80mm}{c@{\extracolsep{\fill}}c}
%\toprule configuration & $C_{x}$ \\
%\hline
%0000 & 20$C$ \\
%0001 & 22$C$ \\
%0010 & 24$C$ \\
%0011 & 26$C$ \\
%0100 & 28$C$ \\
%0101 & 30$C$ \\
%0110 & 32$C$ \\
%0111 & 34$C$ \\
%1000 & 36$C$ \\
%1001 & 38$C$ \\
%1010 & 40$C$ \\
%1011 & 42$C$ \\
%1100 & 44$C$ \\
%1101 & 46$C$ \\
%1110 & 48$C$ \\
%1111 & 50$C$ \\
%\bottomrule
%\end{tabular*}
%\end{center}

\subsection{Digital comparator}

In order to generate veto signals for the other pixel detectors at the EuXFEL, a digital comparator has been implemented in each channel. It is fully synthesized using Cadence tools (rc and encounter)~\cite{Cadence}. It compares the 14-bit output (12 bits from the ADC and 2 bits for coding the gain used by the dynamic gain switching PRE) with a 14-bit threshold which is set by the user. In case the signal induced by X-ray photons is above the given threshold the digital comparator outputs '1' otherwise '0'. This gives the information whether a photon signal above threshold is observed for each channel. The comparator output from every 16 channels of the ASIC is stored into a 16-bit shift register and available for reading out within a time window of $\sim$ 160 ns for each image, which requires a readout clock greater than 100 MHz.

The photon hit information from each channel will be collected and further processed inside the FPGA on the readout board of the Gotthard-II detector, and serves as a basis of the veto signal for the other pixel detectors.

\subsection{Static Random-Access Memory}

%The SRAM implemented into the Gotthard-II readout ASIC is based on the UMC-110 nm SP/AE logic process. It supports synchronous read and write operations. 
The SRAM implemented into the Gotthard-II readout ASIC is a Faraday IP block synthesized with their proprietary memory compiler using standard performance transistors~\cite{Faraday}. It supports synchronous read and write operations.  

16 SRAMs of 2720 memory depth are implemented into the readout ASIC in order to record all the 2700 images from every bunch train of the EuXFEL. The memory address is generated through a 12-bit counter, which starts from 0 and increments by 1 after each image. Every memory address is able to store 112 bits data (also defined as 1 "word") grouped in 8 strip channels, 14 bits each with 12-bit raw output from the ADC as well as 2-bit for gain encoding. 

Before writing into the SRAM, data from the ADC output needs to be stored in a temporary memory consisting of a 112-bit register until all the 112 bits are available due to the pipeline processing. During reading out, data saved in the SRAM is copied to a temporary memory and then shifted out through shift registers. The serial outputs of the shift registers of 2 memory blocks are connected to a 2:1 demultiplexer and then to a physical output pad.%Every two of the long shift registers are connected to a multiplexer and then to a physical output pin of the ASIC.

The usage of the SRAM is mainly for the burst operation at the EuXFEL. In case of CW operation of the Gotthard-II detector, the SRAM is bypassed. Data stored in the temporary memory at the input of the SRAM is directly copied to the temporary memory at the output of the SRAM and then being read out.

\subsection{Periphery}

In the periphery, there are several Digital-to-Analog Converters (DACs), which convert digital inputs to analogue voltages. These reference voltages are used by the comparator, CDS amplifier in the analogue front-end as well as the ADC comparator. We are using 10-bit DACs featuring a resolution of better than 1.5 mV. %The DACs are 10-bit DACs with a resolution of better than 1.5 mV. 

In addition, low output impedance buffers are implemented as well to drive the DAC voltages from the periphery to the front-end. 

\subsection{Layout and geometry}

\begin{figure}[hbt!]
\small
\centering
\includegraphics[width=12cm]{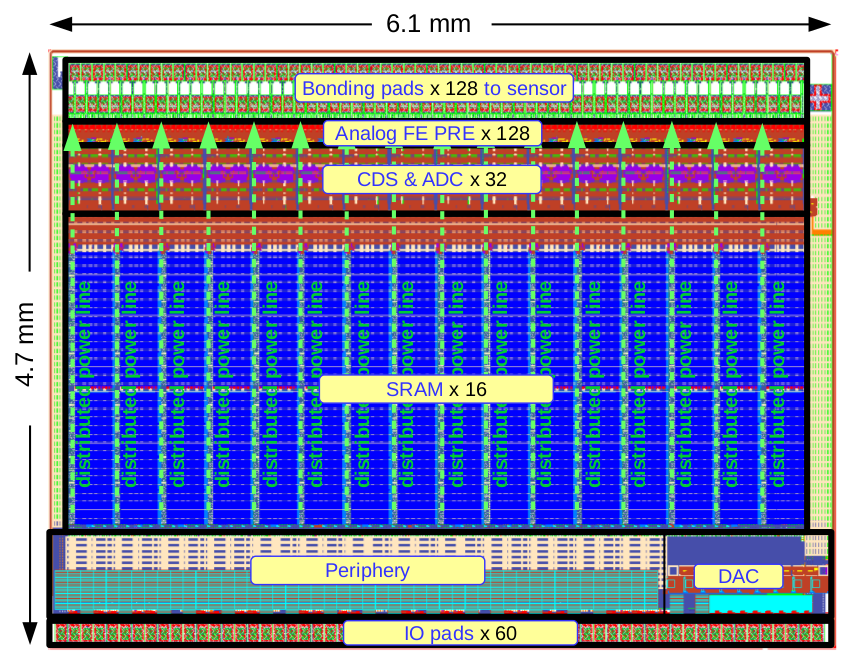}
\caption{The layout and geometry of the Gotthard-II readout ASIC. }
\label{ASIC_layout} 
\end{figure}

The layout of the Gotthard-II readout ASIC is shown in figure~\ref{ASIC_layout}. The width of the readout ASIC is 6.1 mm and the length 4.7 mm. It includes 128 analogue front-end pre-amplifiers, 32 CDS amplifiers and ADCs, 16 SRAMs with a depth of 2720 images, 128 bonding pads for wire-bonding to the sensor, as well as 60 IO pads for ASIC powering, configuration and data output. Within the 60 IO pads, 8 of them are digital output of either image data or hit information for veto generation.

The total power consumption of the Gotthard-II ASIC is less than 300 mW at an operation voltage of \mbox{1.4 V}, corresponding to \mbox{2.3 mW} per channel on average.

\section{ADC calibration}

Due to the parasitic capacitance of the bridge capacitor in the DAC array of the ADC as well as the capacitance mismatch in the DAC array of the ADC, the ADC output may not be linearly dependent on the amplitude of its input signal and thus an ADC calibration is essential as the first step.

\subsection{Calibration procedures}

The ADC calibration in this work relies on a histogram test method. It requires a large amount of data, good statistics as well as a higher precision from the input signal. In the calibration, a linearly increasing voltage is applied to $V_{ref,gx}$ from a wave-generator using a triangular wave to sweep the output of the CDS stage (see formula~\ref{eq2}). The pre-amplifier is fixed to the low gain. The differential voltages at the output of the CDS stage cover the entire input range of the ADC. The generated differential signal at the output of the CDS needs to meet the following requirements as well: 1) good resolution with a voltage step well below 1 Least Significant Bit (LSB) of the 12-bit ADC, 2) good linearity over the entire input voltage range of the ADC if a linear ramp is used, and 3) preferably small or negligible noise compared to the ADC itself.

\begin{figure}[hbt!]
\small
\centering
\includegraphics[width=14.5cm]{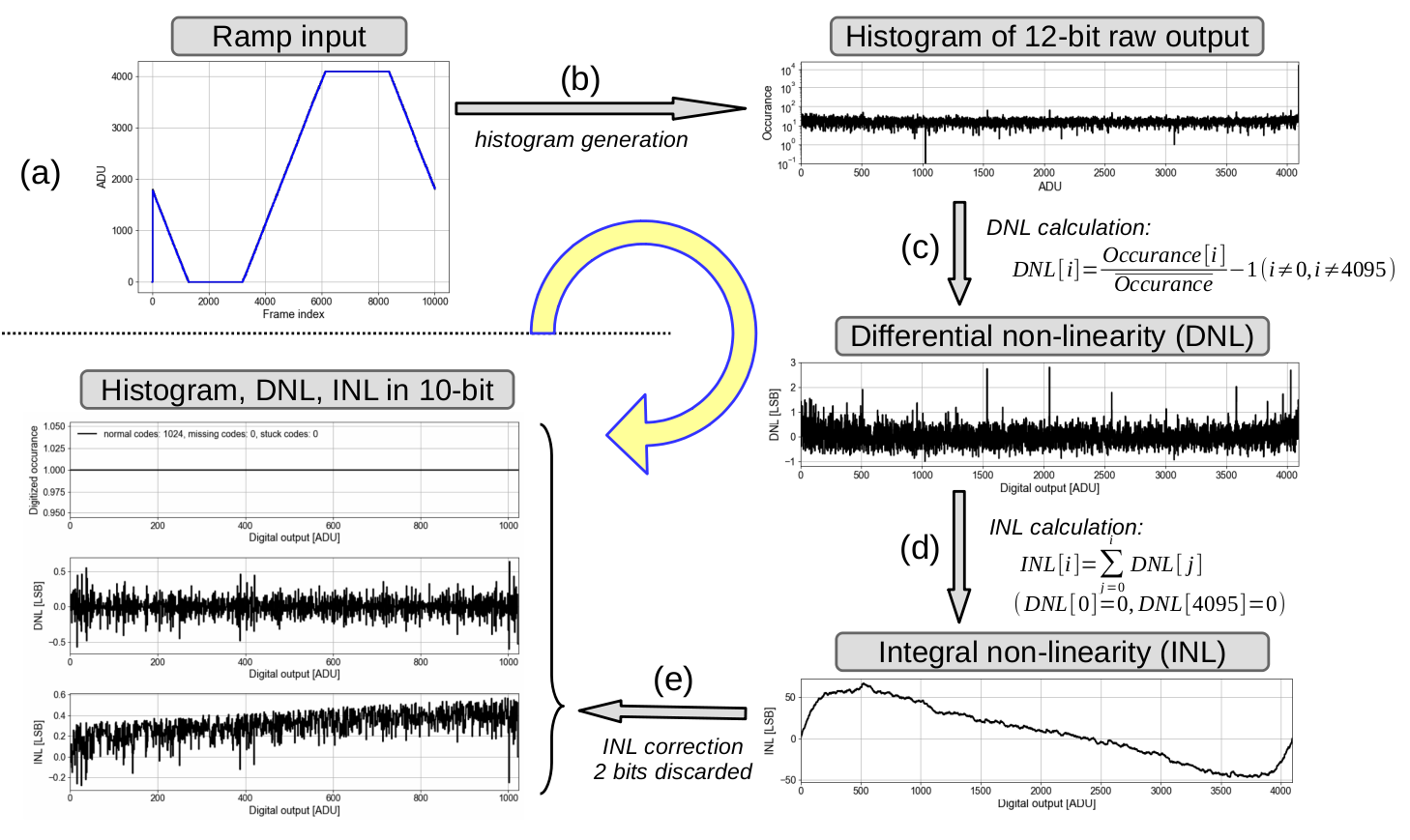}
\caption{Procedures used for ADC calibration to 10-bit. }
\label{ADC_cal} 
\end{figure}

The procedure used for ADC calibration is summarized below and shown in figure~\ref{ADC_cal}:

(a) feed a slow linear voltage ramp generated by the wave-generator to $V_{ref,gx}$ and take a certain number of frames;

(b) generate a histogram from the 12-bit raw output of each ADC;

(c) extract the Differential Non-Linearity (DNL) from the histogram according to:

\begin{equation}
DNL[i] = \frac{N[i]}{\overline{N}}-1
\label{a1}
\end{equation}
\begin{equation*}
(i\neq0,4095)
\end{equation*}

\noindent with $N[i]$ the occurrence of code-$i$ and $\overline{N}$ the average of the occurrence for all codes. If $DNL[i] = -1$, code-$i$ never appears at the ADC output and it is considered as a missing code; if $DNL[i] \geq 1$, code-$i$ appears more often than normal, which covers a wider range for input signal, and thus it is defined as a stuck code.

(d) calculate the Integral Non-Linearity (INL) from the DNL:

\begin{equation}
INL[i] = \sum^i_{j=1} DNL[j]
\label{a2}
\end{equation}
\begin{equation*}
(DNL[0]=0, DNL[4095]=0)
\label{a2}
\end{equation*}

The $INL$ at code-$i$ is defined as the integration of the DNL from 0 to the index of the code. The $INL[i]$ represents the deviation of code-$i$ of the ADC output from a linear output.

(e) correct the 12-bit raw output with the INL and then discard the last 2 bits to achieve 10-bit resolution:

\begin{equation}
Dout_{10bit}[i] = (Dout_{12bit}[i] + INL[i])/4
\label{a3}
\end{equation}

After going through the calibration procedure, a Look Up Table (LUT) is generated for each ADC according to formula~\ref{a3}, from which the 12-bit raw output can be converted to 10-bit calibrated output. In addition to the LUT, the DNL and INL are calculated for the 10-bit output again and they are useful for correcting the spectroscopic performance, which will be discussed later.

\subsection{ADC speed}

A delay can be configured with 3 bits for the comparator of the ADC. The last two bits allow a delay before each comparison decision is made, while the first one only delays the comparisons when the big capacitance in the DAC array, $b_{4}$, $b_{9}$ and $b_{10}$ in figure~\ref{ADC_DAC}, are connected. Increasing the delay, on one hand, improves the signal settling at the input of the comparator and thus results in a better linearity as well as more accurate output; on the other hand, it reduces the sampling/comparison rate of the ADC. For Gotthard-II, it is important to find out a proper setting which satisfies the speed requirement and provides a good linearity at the same time.

The sampling/comparison rate can be measured using an external clock. When the frequency of the external clock is higher than the sampling/comparison rate of the ADC, the ADC stops working and does not send out the $"dout\_rdy"$ signal, which is a signal sent out by the ADC if 12 comparisons can be made before the next sampling/comparison clock arrives. The measurement has been done at different delay configurations. Table~\ref{ADC_speed} shows the measured maximal sampling/comparison rate compared to the post-layout simulation.

\begin{table}[ht]
\centering
\vspace{3.5mm}
\begin{tabular}{c c c}
\hline \hline
configuration & post-layout simulation   & measurement \\ [0.5ex]
\hline
000 & 28 MS/s & (20, 25) MS/s \\
001 & 24 MS/s & (20, 25) MS/s \\
010 & 20 MS/s & (20, 25) MS/s \\
011 & 9.7 MS/s & (9, 10) MS/s \\
100 & 23 MS/s & (20, 25) MS/s \\
101 & 19.6 MS/s & (14, 20) MS/s \\
110 & 17.5 MS/s & (14, 20) MS/s \\
111 & 9.0 MS/s & (9, 10) MS/s \\
[1ex]
\hline
\end{tabular}
\caption{Maximal sampling/comparison rate at different delay settings of the ADC.}
\label{ADC_speed}  
\end{table}

\begin{figure}
\small
\centering
\includegraphics[width=14.5cm]{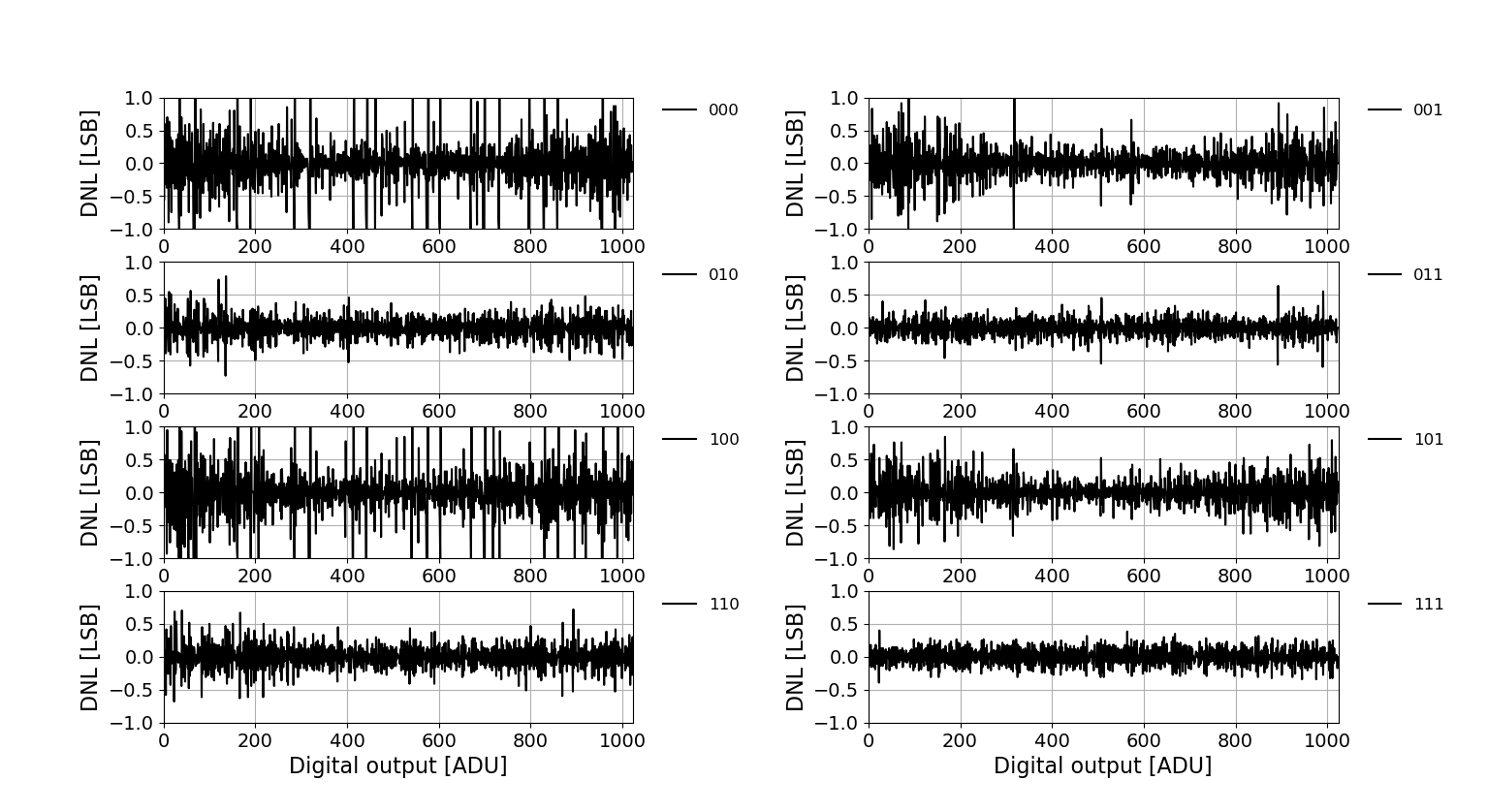}
\caption{The DNL of the ADC after calibration to 10-bit resolution using different delay configurations. }
\label{ADC_speed_DNL} 
\end{figure}

The measurement results, showing a range in which the ADC starts not to work, agree with the post-layout simulation reasonably well. The maximal rate obtained from measurements is \mbox{$\sim$ 25 MHz}, and the minimal \mbox{$\sim$ 9 MHz}. Figure~\ref{ADC_speed_DNL} shows the DNL of the ADC after calibration to 10-bit resolution for different delay configurations. The parasitic capacitance $C_{x}$ was configured to "0100" giving a capacitance value of 28$C$ (28 times the unit capacitance $C$). At higher speed ("000", "001" and "100"), either missing codes ($DNL[i] = -1$) or stuck codes ($DNL[i] \geq 1$), or a combination of both can be observed. These settings are not optimal for the EuXFEL applications using Gotthard-II, which poses a minimal requirement on the sampling/comparison rate of $\geq$ 18 MS/s. Considering the variation of speed between ADCs, the configuration with "010" is suggested which meets the speed requirement as well as leaves enough time for signal settling at the input of the ADC comparator compared to the other settings with higher speed. This setting shows good DNLs at all 10-bit codes (most of them are within \mbox{$\pm$ 0.5 LSB}); in addition, neither missing codes nor stuck codes can be found from the measurement. As shown in figure~\ref{ADC_speed_DNL}, a better data quality using configurations either "111" or "011" for continuous mode at a lower frequency is expected compared to the settings for burst mode at a frate of 4.5 MHz.  %When using Gotthard-II in continuous mode, the configurations either "111" or "011" with minimal sampling/comparison rate are suggested to be used, which gives the best performance of the ADC and thus better data quality if a frame as high as \mbox{$\sim$ 4.5 MHz} is not pursued.

\subsection{ADC side capacitance}

The side capacitance, $C_{x}$, in the DAC array of the ADC can be tuned in order to reduce the influence due to the parasitic capacitance of $C_{bx}$ as well as the the capacitance mismatch of the capacitors in the DAC array. The value of $C_{x}$ can be configured with 4 bits, covering from 20$C$ up to 50$C$ with a difference of 2$C$ per step. 

\begin{figure}
\small
\centering
(a)\includegraphics[width=7.1cm]{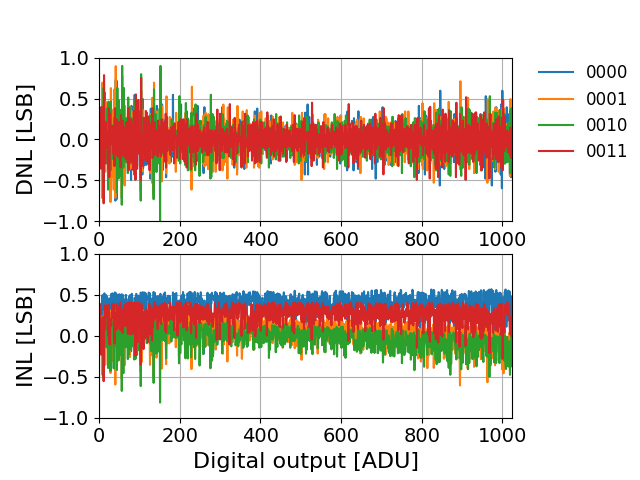}
(b)\includegraphics[width=7.1cm]{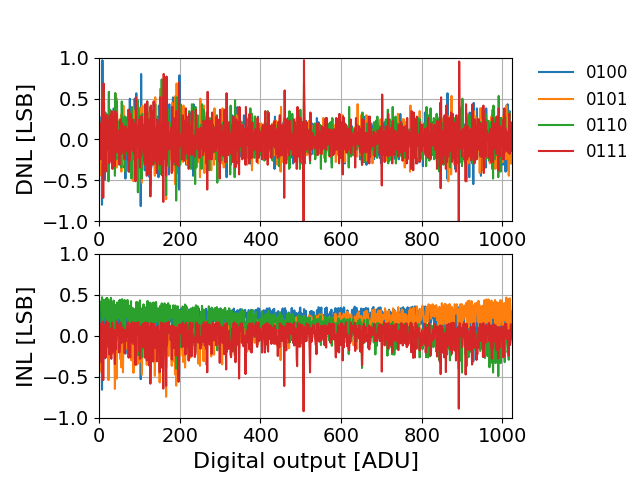}
(c)\includegraphics[width=7.1cm]{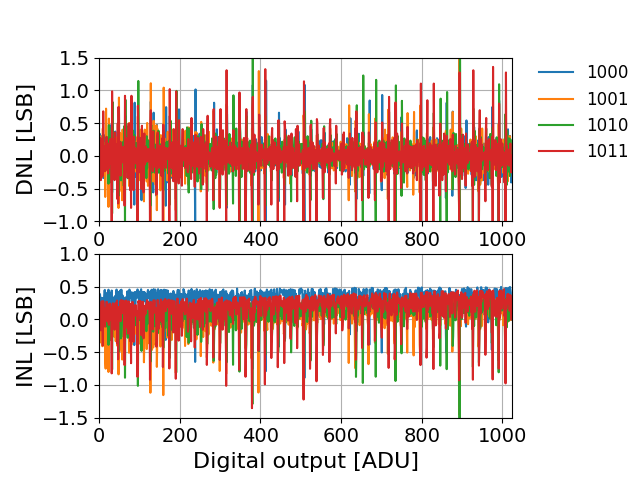}
(d)\includegraphics[width=7.1cm]{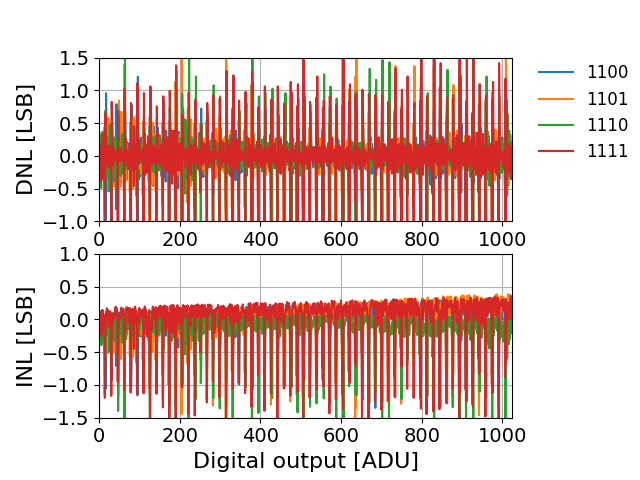}
\caption{The DNL and INL of the ADC after calibration to 10-bit resolution using different side capacitance: (a) for side capacitance configurations "0000", "0001", "0010" and "0011", (b) for "0100", "0101", "0110" and "0111", (c) for "1000", "1001", "1010" and "1011", (d) for "1100", "1101", "1110" and "1111".}
\label{ADCs_parasitic_DNL} 
\end{figure}

Figure~\ref{ADCs_parasitic_DNL} shows the DNL and INL of the ADC after calibration to 10-bit resolution using different side capacitance by tuning the 4 configurable bits. The ADC speed was configured to "010" determined from the previous section. With smaller side capacitance ("0000" to "0110"), the DNL are within $\pm$0.5 LSB for most of the 10-bit output codes and INL within $\pm$1.0 LSB. For bigger side capacitance ("0111" to "1111"), more missing codes ($DNL[i] = -1$) and stuck codes ($DNL[i] \geq 1$) are observed. Since missing codes and in particular stuck codes are difficult to correct, configurations using smaller side capacitance (no missing and stuck codes) is suggested.
%Since it is easier to make corrections for the missing codes, smaller side capacitance for Gotthard-II is suggested to be used.

\subsection{From 12-bit raw output to 10-bit resolution}

\begin{figure}
\small
\centering
\includegraphics[width=14.5cm]{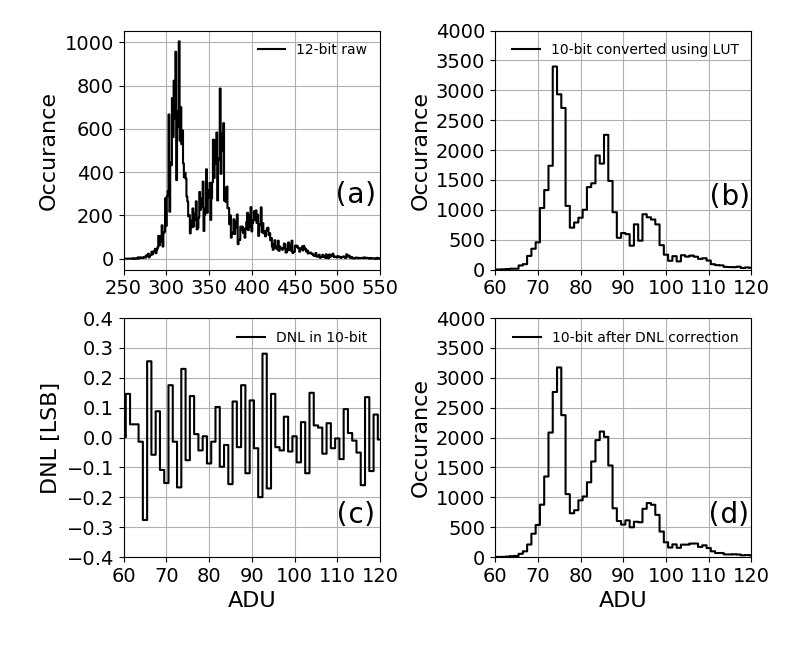}
\caption{(a) Histogram of the 12-bit raw output from a single strip channel of Gotthard-II. (b) Histogram in 10-bit with the 12-bit output calibrated to 10-bit resolution using a LUT. (c) DNL at different codes at 10-bit resolution. (d) Histogram of 10-bit codes after correcting the DNL for each code.}
\label{Spectroscopic_correction} 
\end{figure}

The 12-bit raw output of each ADC can be converted to the 10-bit calibrated data using the LUT extracted following the procedures described in section-3.1. Figure~\ref{Spectroscopic_correction}(a) and (b) show the pulse height distribution of a strip channel when using X-ray fluorescence of 8.05 keV at 12-bit and 10-bit resolution. Such a spectrum is of importance since it is normally used to determine the conversion gain of X-ray photons of a certain energy and further served as a basis of detector calibration for the other two gains after dynamic gain switching.

The spectrum at 10-bit resolution shows large fluctuations. The fluctuation between neighbouring output codes is larger than the uncertainty of the statistic given by the square root of the occurrence at the code. This is mainly caused by the DNL of each code, as shown in figure~\ref{Spectroscopic_correction}(c). For example, the DNL at codes 92 and 94 are about $\sim$ -0.2 and -0.17 respectively, indicating the codes covering narrower range (bin) for input signal than normal and thus occured less; on the contrary, the DNL at codes 93 and 95 are $\sim$ 0.28 and 0.15 indicating wider range coverage of the input signal. The difference in DNL explains why codes 92 and 94 in the histogram in figure~\ref{Spectroscopic_correction}(b) show much lower occurance compared to codes 93 and 95. The non-uniform response of each code at 10-bit caused by the difference in DNL can be corrected by:

\begin{equation}
N_{10bit,corr}[i] = \frac{N_{10bit}[i]}{DNL_{10bit}[i]+1}
\label{a4}
\end{equation}

\noindent according to formula~\ref{a1}.

Figure~\ref{Spectroscopic_correction}(d) shows the pulse height distribution after correcting DNL using formula~\ref{a4}. The response of each code becomes consistent. This method in principle can be applied to spectroscopic experiments as well when the spectrum distribution from individual strip channels becomes important. The DNL at 10-bit resolution is not expected to take an impact on the performance of Gotthard-II in energy dispersive experiments at the EuXFEL, where the INL is more important. However, for detector calibration, in particular the extraction of conversion gain, the influence of the DNL has to be taken into account.

\section{Characterization of the Gotthard-II ASIC}

The Gotthard-II ASIC has been wire-bonded to a \mbox{320 $\mu$m} thick silicon microstrip sensor for the characterization. The pitch of the strip is 50 $\mu$m and the length is 8 mm. During the tests, the sensor was biased at \mbox{120 V}. 

For all the tests discussed below, the ADC speed was configured to "010" and the side capacitance to "0100", which are considered as the default setting for the EuXFEL applications. 

\subsection{Single-photon sensitivity}
\label{sec:noise}

In order to test the single-photon sensitivity, X-ray fluorescence of 8.05 keV from a Copper target was used. The Gotthard-II ASIC was configured in burst mode with frames taken from X-ray fluorescence stored into the on-chip SRAM. An exposure time of 5 $\mu$s was used to measure a reasonable amount of X-ray photons per frame per strip channel.

\begin{figure}[hbt!]
\small
\centering
(a)\includegraphics[width=7.1cm]{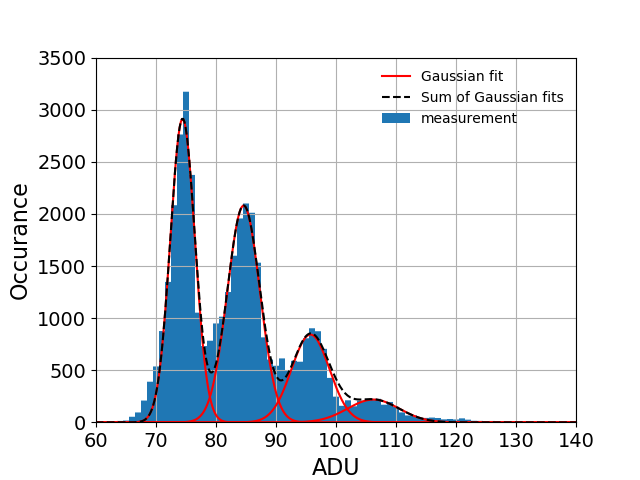}
(b)\includegraphics[width=7.1cm]{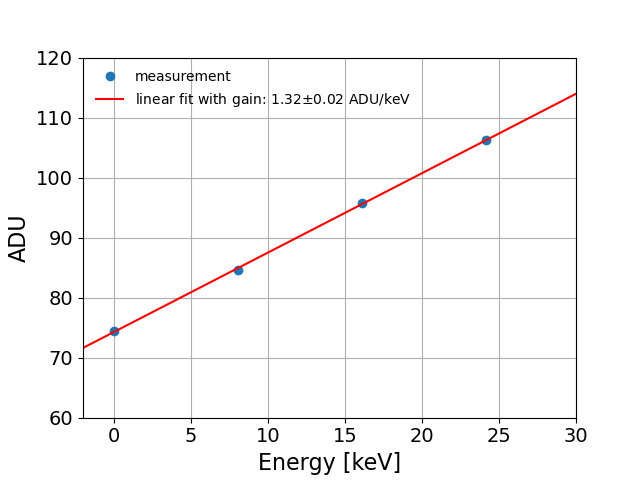}
\caption{(a) Histogram of the output from a single strip channel after calibration to 10-bit and DNL correction, which shows single-photon resolution for 8.05 keV X-ray fluorescence at low flux. (b) Extracted peak position from a Gaussian fit to each individual peak as function of energy for the determination of the conversion gain. }
\label{Single_photon_resolution} 
\end{figure}

Figure~\ref{Single_photon_resolution}(a) shows the histogram of the Gotthard-II output from a single strip channel. Four peaks can be identified from the measurement: The first peak refers to no photon hits within the exposure time, the second peak to the hits of a single photon of 8.05 keV, as well as the third and the fourth to the hits of 2 or 3 coincidence X-ray photons within the exposure time. The measurement shows single photon resolution at 8.05 keV before gain switching.

The peak positions of these identified peaks can be extracted through a Gaussian fit to each individual peak. Figure~\ref{Single_photon_resolution}(b) shows the extracted peak position in ADU as a function of energy and a linear fit to the experimental data. The conversion gain given by the slope of the linear fit is $g_{0}=1.32\pm0.02\ ADU/keV$. 

From the Gaussian fit to the first peak in figure~\ref{Single_photon_resolution}(a), a sigma of 1.3 ADU can be obtained indicating an electronic noise of $\sim$ 273.6 $e^{-}$, which is calculated by $noise\ r.m.s. = \sigma /g_{0} \cdot 1000/3.6$. Such a noise indicates a signal-to-noise ratio (SNR) greater than 10 for 12.4 keV X-rays. The SNR provided refers to the separation between the single photon and the zero photon peak. %5 for 5 keV X-rays and 10 for 10 keV X-rays with excellent single photon resolution for hard X-rays above 5 keV.

\subsection{Dynamic range, non-linearity and noise}

The dynamic range was measured using the internal current injection circuitry of the ASIC. The current was injected into the input of the pre-amplifier during the signal sampling phase when the PRE output is connected to $C_{s}^{E/O}$. Assuming the injected current is stable over time, by varying the exposure time during current injection, the total amount of charge stored on the feedback capacitor of the pre-amplifier increases linearly with the exposure time at each gain. With the current injection, the Gotthard-II output is measured as a function of exposure time. The 12-bit raw output of the ADC has been converted to 10-bit calibrated output using the LUT. The exposure time can be calibrated to the number of 12.4 keV photons by comparing the slope in $G0$ in units of ADU/ns to the extracted conversion gain in ADU/keV, which gives the factor needed to scale the exposure time to photon energy.

\begin{figure}[hbt!]
\small
\centering
(a)\includegraphics[width=7.1cm]{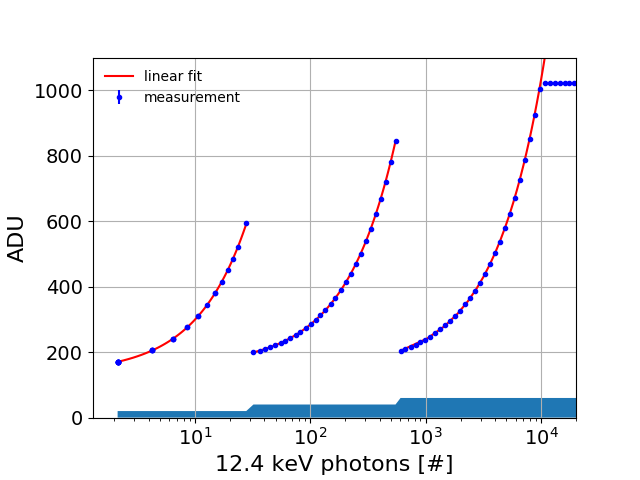}
%(b)\includegraphics[width=7.5cm]{DR_residual.png}
(b)\includegraphics[width=7.1cm]{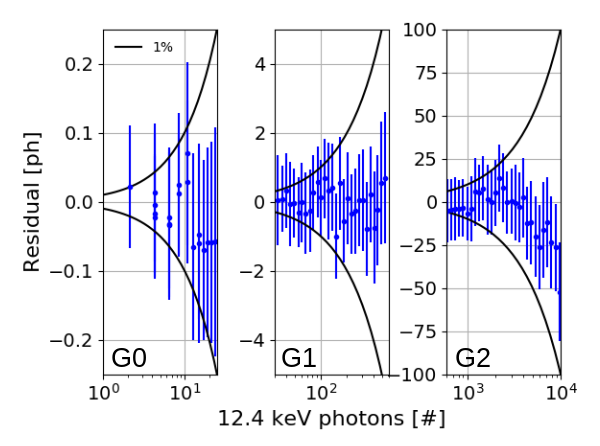}
(c)\includegraphics[width=7.1cm]{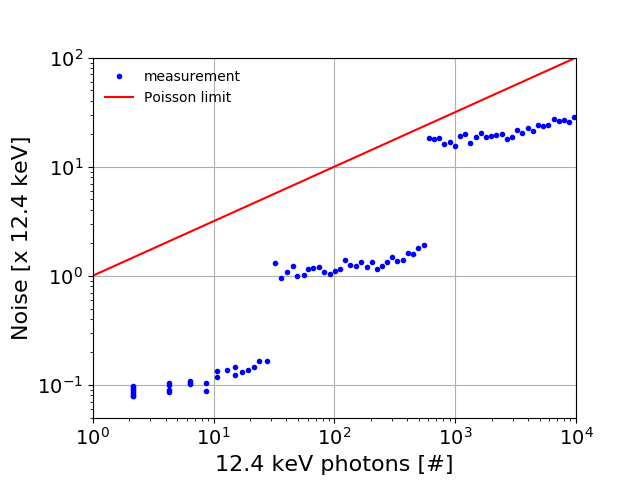}
\caption{(a) Dynamic range of the Gotthard-II ASIC scanned using an internal current source. The horizontal blue lines at the bottom show the 3 gain ranges. (b) Non-linearity of the Gotthard-II output at G0 (left), G1 (middle) and G2 (right). (c) Noise over the entire dynamic range.}
\label{Dynamic_range} 
\end{figure}

Figure~\ref{Dynamic_range}(a) shows the measured curve in the dynamic range scan after converting the exposure time to photon energy expressed in number of 12.4 keV photons: The blue dots are from measurement and the red lines are linear fits to each gain. From the measurement, a dynamic range of up to $10^{4}$ 12.4 keV X-ray photons before saturation can be observed. Due to the pre-charge scheme implemented in the pre-amplifier design, it is capable of extending and maximizing the dynamic range compared to a similar design without pre-charge, $e.g.$ AGIPD with a dynamic range of \mbox{$\sim$ 7000 $\times$ 12.4 keV} X-ray photons~\cite{Davide2016, Davide2019, Ashkan2014}. From the linear fit, the conversion gain in $G0$, $G1$ and $G2$ can be obtained, which corresponds to \mbox{1.32 ADU/keV}, \mbox{0.10 ADU/keV} and \mbox{0.007 ADU/keV}, respectively. The gain ratios between $G0$ and $G1$, as well as $G1$ and $G2$ are 13.2 and 14.3. These ratios agree well with the design values (differences within 10\%).

The residuals in units of 12.4 keV photons are calculated using the difference between the measurement and the linear fit divided by the conversion gain of $G0$, $G1$ and $G2$. The result is shown in figure~\ref{Dynamic_range}(b): The non-linearity of the detector output is less than 1\%, showing excellent linearity over the entire dynamic range up to $10^{4}$ 12.4 keV photons.

The variation at each exposure time in the measurement is also extracted. Since it is mainly caused by the electronic noise of Gotthard-II, the variation has been converted to noise using the determined gains of $G0$, $G1$ and $G2$. Figure~\ref{Dynamic_range}(c) shows the noise in units of 12.4 keV photons over the entire dynamic range. In the different gains, the noise values are equivalent to $\sim$ 0.1, 1.3 and 19.0 $\times$ 12.4 keV photons, which are well below the Poisson limit of X-ray photons shown in red line (square root of number of incoming photons). 

%It should be noted that at the end of each gain, the noise increases by a factor of 2, which is caused by the fluctuation of charge injection by the current source.

\subsection{Frame rate and SRAM tests}

In order to verify the frame rate and the storage capability of the SRAM, $V_{ref,gx}$ was connected to a wave-generator, which generated either a triangle or a sine waveform with a period of \mbox{100 $\mu$s}. In addition, the internal current circuitry was used to generate a charge which is high enough to cause gain switching so that $V_{ref,gx}$ is connected to the input of the CDS stage and thus the detector output varies according to the value of $V_{ref,gx}$ fed by the wave-generator. 

\begin{figure}[hbt!]
\small
\centering
\includegraphics[width=10cm]{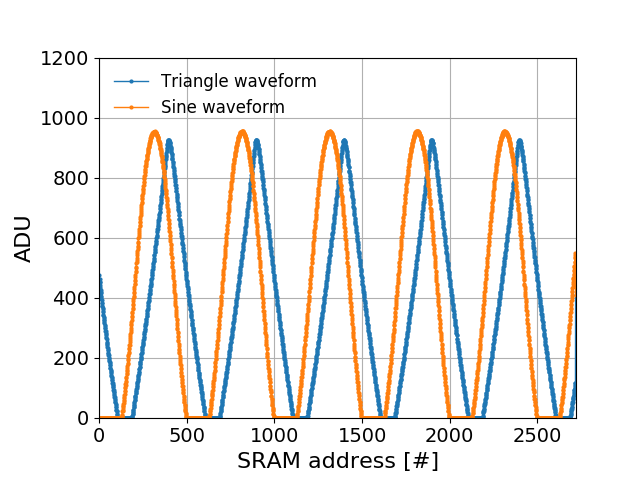}
\caption{Test of the SRAM with 2720 memory addresses at 5 MHz frame rate.}
\label{SRAM_test} 
\end{figure}

In the measurement, a frame rate of 5 MHz was used, and the Gotthard-II ASIC was configured in a burst mode with 2720 images stored in the SRAM and then read out later. Figure~\ref{SRAM_test} shows the output from the 2720 memory addresses of the SRAM after calibrating to 10 bits (the first 2 bits coding the gain have been ignored): All data can be properly saved into SRAM at 5 MHz frame rate. It can be seen that the detector output is repeated every 500 memory addresses, which corresponds to 200 ns/address $\times$ 500 addresses = 100 $\mu$s and agrees with the period set from the wave-generator.

%From this simple test, a frame rate of 5 MHz for up to 2720 images can be demonstrated.

From this measurement, a frame rate greater than 4.5 MHz has been demonstrated using the default configuration for the EuXFEL. Since the frame rate is dominated by the ADC speed, when setting the ADC to its highest sampling/comparison rate, namely $\sim$ 25 MS/s using "000" configuration as shown in Table~\ref{ADC_speed}, a frame rate of $\sim$ 6.25 MHz can be considered as the maximal rate of the Gotthard-II ASIC.

\subsection{Signal settling time}

The resistor connected to the output of the PRE, $R_{filt}$, helps to filter the high frequency noise. However, it influences the time required to write the charge into the storage capacitors of the "signal \& reset sampling stage" (see figure~\ref{Architecture}). To measure the settling time, a laser delay scan was performed. An exposure time of 5 $\mu$s was set to Gotthard-II and the ASIC was configured to a frame rate of \mbox{1.125 MHz}. An infrared laser light with a wavelength of 1030 nm was used and triggered by an external signal which can be delayed with respect to the integration window of the detector. The infrared laser injected into the silicon sensor produces charge, which is integrated onto the feedback capacitor of the PRE and thus moves its output. Details of the measurement set-up and principle have been documented elsewhere~\cite{Davide2016}.

\begin{figure}[hbt!]
\small
\centering
\includegraphics[width=10cm]{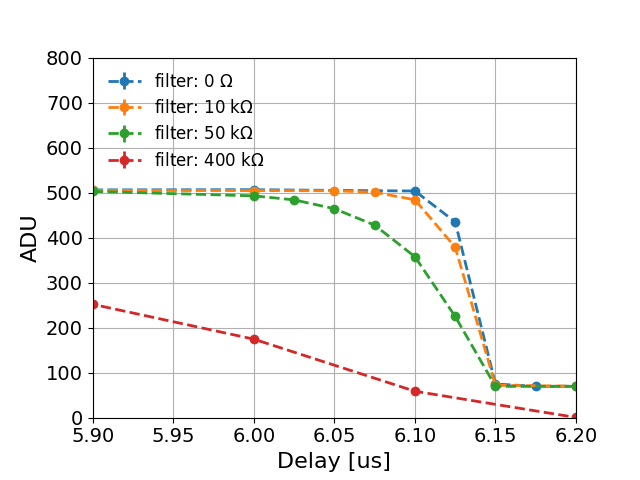}
\caption{Laser delay scan with an exposure time of 5 $\mu$s: The Gotthard-II ASIC output at 10-bit as function of delay time of the laser trigger for different filtering resistance.}
\label{Delay_scan} 
\end{figure}

Figure~\ref{Delay_scan} shows the 10-bit output of a single strip with laser injection as function of delay time for different resistance configured to $R_{filt}$. At the delay of 6.15 $\mu$s, the infrared laser was injected outside the integrating window and thus the output equals to the pedestal for resistance of 0, 10 k$\Omega$ and 50 k$\Omega$. The transition times from the full charge, 500 ADU in figure~\ref{Delay_scan}, to the pedestal refer to the signal settling time. For 400 k$\Omega$ resistance, since the settling time is too long, there is not enough time for $C_{r}^{E/O}$ to sample the "reset" signal before the integration window is open, which results in a serious shift in the pedestal.

From the measurement, it requires $\sim$ 50 ns, $\sim$ 75 ns, $\sim$ 250 ns and $>$ 600 ns for resistance of 0, 10 k$\Omega$, 50 k$\Omega$ and 400 k$\Omega$, respectively. The filtering resistance of 0 and 10 k$\Omega$ can be configured and used in burst mode operation at the EuXFEL: the nominal noise can be improved by $\sim$ 8\% when the 10 k$\Omega$ filter resistor is selected compared to the result shown in section~\ref{sec:noise} with \mbox{0 k$\Omega$} resistance.  %while 50 k$\Omega$ and 400 k$\Omega$ for operation in continuous mode with lower frame rates.

\subsection{Cross-talk effect}

Due to the large interstrip capacitance from the long strips of the silicon sensor, a cross-talk effect is often observed in silicon strip detectors. The cross-talk effect depends on the interstrip capacitance, feedback capacitance of the charge-sensitive PRE, as well as the DC gain of the PRE, which has been described in detail in \cite{Zhang2017, Zhang2018}. For a low DC gain PRE using a dynamic gain switching architecture, when a strip channel switches its gain its neighbouring strip channel can observe a significant reduction in its output. Such a change can be as large as $\sim$ 3.3 $\times$ 12.4 keV X-ray photons, as reported in \cite{Zhang2017}. In this case, conversion of the channel output to photon energy depends on the gain of its neighbouring strip channel and thus it makes calibration of the detector system complex. To solve the problem, a high DC gain PRE with a DC gain greater than 600 has been designed and first implemented in the Gotthard-1.7 analogue front-end prototype \cite{Zhang2018}. After verifying its improvement on the cross-talk, the same design has been used in the final Gotthard-II ASIC. 

%Figure~\ref{Cross_talk} shows the result after converting to 10 bits of dynamic range scan for an injected channel using the internal current source as well as the output of its neighbouring channel. 

\begin{figure}[hbt!]
\small
\centering
\includegraphics[width=10cm]{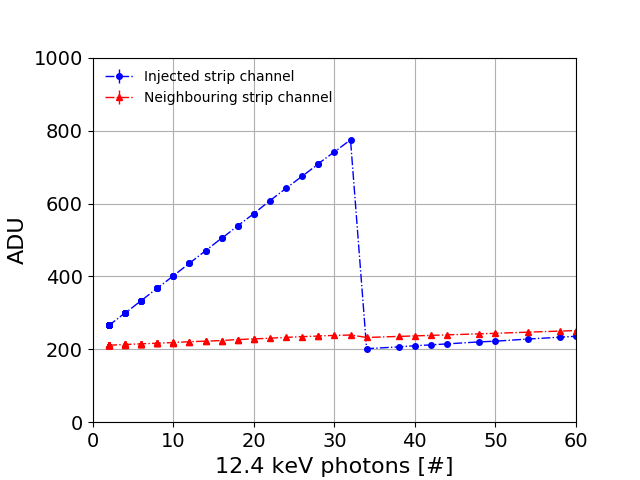}
\caption{Cross-talk at the gain switching point. The region between $G0$ and $G1$ is shown and emphasized.}
\label{Cross_talk} 
\end{figure}

Figure~\ref{Cross_talk} shows the result of dynamic range scan for a channel through current injection using its internal current source as well as the output of its neighbouring channel. The results are shown after converting to 10 bits. Every point in the figure is the result of averaging 100 repeated measurements. The injected strip channel is shown in blue dots while its neighbouring strip channel without injection in red triangles. The gain switching for the injected channel happens at $\sim$ 32 $\times$ 12.4 keV photons. The change of the output of its neighbouring non-injected channel in red triangles is 6.7 ADU after the gain switching of the injected channel. Considering the conversion gain in $G0$ is 1.32 ADU/keV determined from the measurement of X-ray fluorescence, the change of 6.7 ADU corresponds to 5.08 keV or 0.41 $\times$ 12.4 keV X-ray photon. Since the noise for the channel with gain switched to $G1$ is larger than 1 $\times$ 12.4 keV photons as shown in figure~\ref{Dynamic_range}(c), the cross-talk due to the dynamic gain switching in the Gotthard-II final ASIC is negligible when summing up the total charge measured by strip channels with X-ray photons detected. The 0.41 photon of 12.4 keV is not negligible in the neighbouring channel but would not be counted as a photon when quantitizing the number of photons. For lower photon energies, the value is smaller than Poisson statistical fluctuation and thus is negligible as well. The cross-talk effect has been improved significantly compared to a conventional design reported in \cite{Zhang2017}.

\subsection{Photon hit information}

The photon hit information is extracted from the output of the digital comparator of the Gotthard-II readout ASIC. For the tests, infrared laser light with a wavelength of 1030 nm was injected from the strip side of the silicon sensor in order to create a large signal.

\begin{figure}[hbt!]
\small
\centering
(a)\includegraphics[width=7.1cm]{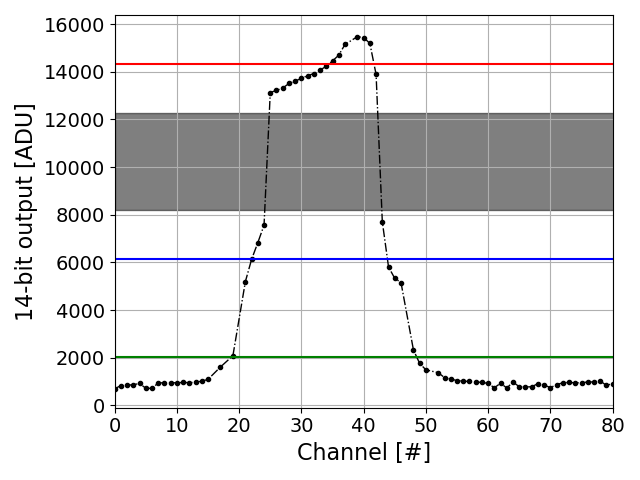}
(b)\includegraphics[width=7.1cm]{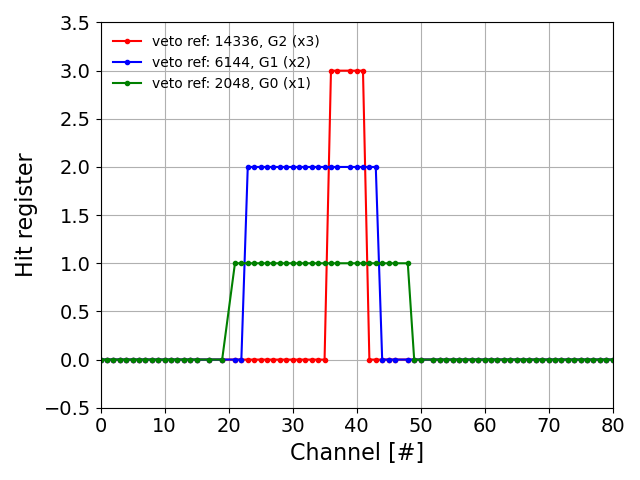}
\caption{(a) The 14-bit output of the Gotthard-II ASIC with laser injection for three veto reference settings of 2048 (green), 6144 (blue) and 14336 (red). (b) The output of the hit shift register for the investigated channels. Note that the outputs for $G1$ and $G2$ are multiplied by 2 and 3 respectively in order to separate the three curves. A mask for the bad channels, 16, 18, 20, 38, 47 and 51 from this assembly has been applied.}
\label{Veto} 
\end{figure}

Figure~\ref{Veto}(a) shows the 14-bit output of the Gotthard-II ASIC consisting of 2 gain bits as well as the ADC raw output of 12 bits. For the digital comparator, the gain bits are coded as "00" for $G0$, "01" for $G1$ and "11" for $G2$, thus codes from 8192 to 12287 (from $2^{13}$ to $2^{13} + 2^{12} -1$) are not available and marked in black. The injected laser light creates signals in different channels covering all three gains. Three different veto references have been set in the tests: 2048 (in $G0$), 6144 (in $G1$) and 14336 (in $G2$). The outputs of the hit register from the investigated channels are shown in figure~\ref{Veto}(b). For veto reference settings of 6144 (in $G1$) and 10240 (in $G2$), the hit register output has been multiplied by 2 and 3 respectively in order to separate the three curves for a better visibility. The channels showing photon hits (hit register is given by 1) in figure~\ref{Veto}(b) agree with those channels with 14-bit output crossing the veto references as shown in figure~\ref{Veto}(a).

The photon hit information serves as a basis for generating veto signals to the other pixel detectors. Since the veto signal generation is done through the FPGA on the readout board of the Gotthard-II detector module, it will be discussed in a separate paper.

\section{First experimental test}

\begin{figure}[hbt!]
\small
\centering
(a)\includegraphics[width=5.5cm]{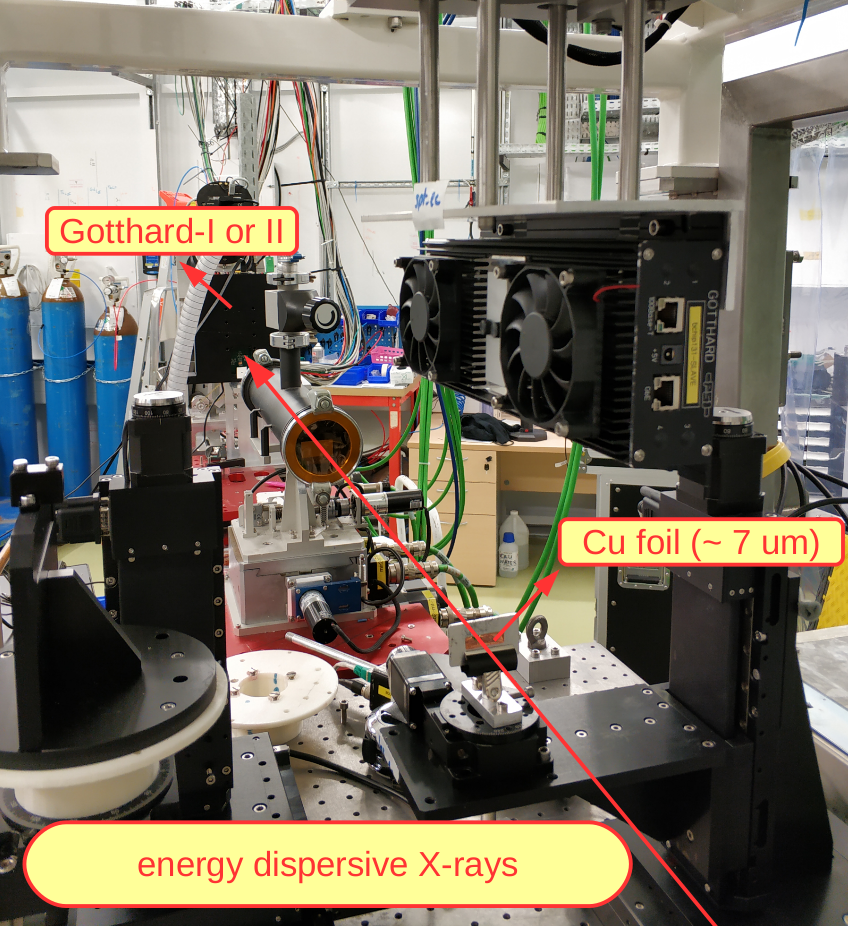}
(b)\includegraphics[width=8.5cm]{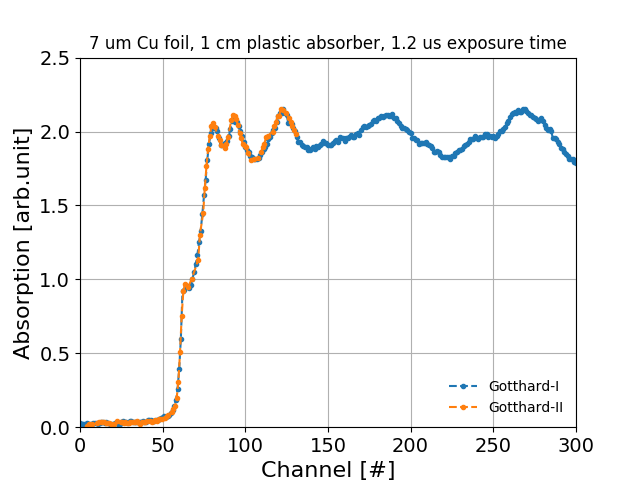}
\caption{(a) The experimental set-up at the ODE beamline of Soleil synchrontron. (b) XAS spectra measured with Gotthard-1 detector and Gotthard-II single chip assembly.}
\label{XAS} 
\end{figure}

As the main 1-dimensional detector for the energy dispersive experiment at the EuXFEL, a first experimental test for the measurement of X-ray Absorption Spectrum (XAS) using a Gotthard-II single chip assembly has been performed. The measurement was done at the ODE beamline of Soleil, which produces energy dispersive X-rays at the end station. The experimental set-up is shown in figure~\ref{XAS}(a). The synchrotron ran in a hybrid mode during the test: it is filled with three cluster of bunches as well as one single isolated bunch. Each cluster with a period of 295 ns carries 104 equal-spaced bunches. The single isolated bunch is filled in the middle of the third and the first bunch cluster, which are separated by 295 ns. Thus one cycle corresponds to $295\ \textrm{ns} \times 4 = 1180\ \textrm{ns}$. 

The dispersive X-rays were aligned to energies from 8.80 keV to 9.12 keV, covering the $k_{\beta}$ edge of Cu. In the measurement, a Cu foil of $\sim$ 7 $\mu$m was placed at the focusing point of the X-ray beam; the detector was about 1 meter away from the Cu foil. To prevent the absorption of X-rays in air, a vacuum pipe was placed in-between the detector and the Cu foil. Since the sensor is exposed to the X-ray beam directly, to avoid radiation damage, an additional plastic absorber of 1 cm thickness was placed in front of the detector. 

An exposure time of $\sim$ 1.2 $\mu$s was set for the Gotthard-II single chip assembly and 1000 frames were taken in total for each measurement. Three measurements were performed: 1) dark measurement for the extraction of the pedestal; 2) $I_{0}$ measurement for X-ray beam without a sample; 3) $I_{1}$ measurement with the 7 $\mu$m thick Cu foil placed at the focusing point of the X-ray beam. For each measurement, the detector output was converted to 10 bits using the LUT before further processing. The absorption for every strip channel is calculated as:

\begin{equation}
absorption \textrm{[arb.unit]} = ln(\frac{I_{0}-pedestal}{I_{1}-pedestal})
\label{absorption}
\end{equation}

The XAS spectrum measured with the Gotthard-II single chip assembly is shown in orange dots in figure~\ref{XAS}(b). The absorption edge at Channel-65 corresponding to a photon energy of \mbox{$\sim$ 8.979 keV} is clearly visible.

In addition, a Gotthard-I detector with 1280 strip channels was used in the same measurements for comparison. The measured XAS spectrum from the Gotthard-I detector is shown in blue dots in figure~\ref{XAS}(b). The absorption edge as well as the first three peaks in the XAS spectrum, which only can be covered by the Gotthard-II single chip assembly due to the limited number of channels, overlap well between the Gotthard-I detector and Gotthard-II single chip assembly. It can be concluded that Gotthard-II shows similarly good XAS performance as Gotthard-I in energy dispersive experiments.

\section{Summary and discussion}

The final Gotthard-II ASIC has been designed and fabricated in \mbox{UMC-110 nm} CMOS technology. The working principle of the Gotthard-II ASIC as well as the procedures for ADC calibration have been described. In addition, a detailed characterization has been performed. The characterization results show a good single-photon resolution at photon energy $>$ 5 keV with a signal-to-noise ratio greater than 5, as well as a large dynamic range up to $10^{4}$ 12.4 keV X-ray photons. For X-ray photons with energy below 5 keV, single-photon resolution cannot be easily achieved due to the limit of electronic noise, which is a compromised result after optimizing frame rate as well as the DC gain of the pre-amplifier. Nevertheless, the noise is below the Poisson limit over the entire dynamic range for 12.4 keV X-ray photons; the non-linearity is better than 1\% over the entire dynamic range as well after calibrating the ADC to 10 bits. In addition, the frame rate has been tested up to 5 MHz for 2720 images to satisfy the requirement at the EuXFEL. All the characterization results obtained so far meet the specifications of the Gotthard-II development. The main parameters of the Gotthard-II ASIC are summarized below in Table~\ref{Main_parameters}.

\begin{table}[ht]
\centering
\vspace{3.5mm}
\begin{tabular}{c c c}
\hline \hline
parameter & value & remark \\ [0.5ex]
\hline
number of strips & 128 & for each ASIC  \\
dynamic range & $10^{4} \times 12.4$ keV & maximal range \\
non-linearity & < 1\% & over the entire dynamic range \\
noise & 273.6 $e^{-}$ & in $G0$ \\
SNR & $>$ 10 & for 12.4 keV photons \\
single-photon resolution & > 5 keV & in $G0$, with SNR better than 5 \\
frame rate & 4.5 MHz & burst mode, max. 6.25 MHz \\
ADC resolution & 10 bits & after calibration \\
ADC speed & 18 MS/s & tunable to max. 25 MS/s \\
SRAM depth & 2720 & maximal images \\
[1ex]
\hline
\end{tabular}
\caption{The main parameters of the Gotthard-II ASIC.}
\label{Main_parameters}  
\end{table}

In addition, one experimental test was performed to measure the X-ray absorption spectrum from a Cu foil using a Gotthard-II single chip assembly. The performance is as good as that obtained using the Gotthard-I detector. Therefore, the performance of Gotthard-II in energy dispersive experiments is demonstrated at least in a synchrotron radiation environment.

\section*{Acknowledgments}

The first experimental test using Gotthard-II was done at the ODE beamline of Soleil. J. Zhang would like to thank the beamline staff, Q. Kong, X. Liu and C. Hua, for the help with tuning the X-ray energy, aligning the X-ray beam as well as preparing the sample used in the experiment, as well as to T. Weng from ShanghaiTech University for the support of the spectrometer.

%\acknowledgments

%This is the most common positions for acknowledgments. A macro is available to maintain the same layout and spelling of the heading.

%\paragraph{Note added.} This is also a good position for notes added after the paper has been written.


\end{document}